\documentclass[twocolumn,showpacs,aps,amsmath,pre]{revtex4}

\usepackage{graphics}
\usepackage{psfig,epsfig}
\usepackage{pstricks}
\usepackage{amsmath}
\usepackage{amsfonts}
\usepackage{xspace}

\newcommand{\GC}{\mathcal{G}}
\newcommand{\gc}{g}

\newcommand{\ave}[1]{\left\langle#1 \right\rangle}

\newcommand{\prepsection}[1]{}

\newcommand{\elabel}[1]{\label{eq:#1}}

\newcommand{\Eref}[1]{Eq.~(\ref{eq:#1})}

\newcommand{\Fref}[1]{Fig.~\ref{F:#1}}

\begin{document}
\title{Avalanche Behavior in an Absorbing State Oslo Model}
\author{Kim Christensen}
\email{k.christensen@imperial.ac.uk}
\author{Nicholas R. Moloney}
\email{n.moloney@imperial.ac.uk}
\author{Ole Peters}
\email{ole.peters@imperial.ac.uk}
\affiliation{Physics of Geological Processes, University of Oslo, PO Box 1048, Blindern,
N-0316 Oslo, Norway \\
Permanent: Blackett Laboratory, Imperial College London, Prince 
Consort Road, London SW7 2AZ, United Kingdom}
\author{Gunnar Pruessner}
\email{gunnar.pruessner@physics.org}
\affiliation{Physics Department, Virginia Polytechnic Inst. \& State
Univ., Blacksburg, VA 24061-0435, USA}

\date{\today}
\begin{abstract}
Self-organized criticality can be translated into the language of
absorbing state phase transitions. Most models for which this analogy is
established have been investigated for their absorbing state
characteristics. In this article, we transform the self-organized
critical Oslo model into an absorbing state Oslo model and analyze the
avalanche behavior. We find that the resulting gap exponent, $D$, is
consistent with its value in the self-organized critical model. For the
avalanche size exponent, $\tau$, an analysis of the effect of the
external drive and the boundary conditions is required.
\end{abstract}
\pacs{89.75.-k, 
      89.75.Da, 
      05.65.+b, 
      45.70.Ht  
      }
\maketitle

\prepsection{Introduction} Self-organized criticality (SOC) refers to
the tendency of non-equilibrium dissipative systems, with many degrees
of freedom driven at a slow rate, to display scale invariance without
the fine-tuning of any control parameter \cite{Bak87,Jensen:98}.  SOC is
present in many \emph{open} systems where the activity, that is, the
presence of avalanches, transports slope particles through the system 
to the boundaries where they are eventually dissipated. When there is no
activity, the system is in a so-called absorbing state.  The separation
of relaxational and driving timescales is achieved by adding a slope unit
only when the system is in an absorbing state. Thus, the dynamics of the
model implicitly tune the slope density to values that are associated
with the transition between absorbing (inactive) and active states.

In a \emph{closed} system, an absorbing state (AS) phase transition
refers to the transition from an absorbing (inactive) state to an active
state of the system at a critical value of a control parameter such as
the slope density \cite{Hinrichsen:2000}.

In retrospect, Tang and Bak's 1988 description of self-organized
criticality contains the ideas and features of absorbing
state phase transitions \cite{Tang88b}.  This link was later clarified
by Vespignani and Zapperi \cite{Vespignani97} and a recipe for
transforming AS models into equivalent SOC models was devised
\cite{Dickman99}: Start with a system displaying an AS
phase transition. When the control parameter is above its critical value,
there is activity in the system. This activity should be coupled via the
dynamics to a decrease in the control parameter, as for example when the
activity reaches the boundary.  Conversely, when the control parameter is
below its critical value, there is no activity and the system is in an absorbing state.
A process, such as the external drive, should increase the control parameter by a small amount
which may force the system into an active state.  However, this picture ignores, for example, the
problem of defining observables common to both AS and SOC models as well
as the r\^ole of finite-size effects.

In the following, we transform the SOC Oslo model with one \emph{open}
boundary into an AS Oslo model with \emph{periodic} boundary
conditions. The difficulties in this procedure are shortly
discussed. Extensive numerical simulations are analyzed with respect to
scale invariance. The results lead to a discussion on the
effect of the external drive and the boundary conditions.

\prepsection{The SOC Oslo model} The model \cite{ChristensenETAL:1996} was
inspired by an experiment, conducted in Oslo, on slowly driven
rice-piles displaying self-organized criticality \cite{Frette96}.  A
one-dimensional lattice of length $L$ is characterized by a slope
variable, $z_i$, and a critical slope, $z_i^c$, assigned to each site
$i=1,\dots,L$. After initialization with $z_i=0$ and $z_i^c$ drawn
randomly with equal probability from $\{1,2\}$, the model is updated as
follows: \emph{Driving}: A slope unit is added to the leftmost site
$i=1$, such that $z_1\to z_1+1$. \emph{Toppling}: If $z_i>z_i^c$ at a
site $i$, one slope unit is moved to each of the two nearest neighbors,
that is, $z_i\to z_i-2$ and $z_{i\pm1}\to z_{i\pm1}+1$ except when site
$i=1$ topples, where $z_1 \to z_1 - 2$ and $z_2 \to z_2 + 1$ or when
site $i=L$ topples, where $z_L \to z_L-1$ and $z_{L-1}\to z_{L-1}+1$.
After each toppling a new value for $z_i^c$ is chosen randomly with
equal probability from $\{1,2\}$. The activity stops when $z_i\le z_i^c$
everywhere. The model is then driven again.

After a transient, the slope density, $\zeta = \left( 1/L \right) \sum_{i=1}^L z_i$,
fluctuates about a constant value. The avalanche size, $s$, is the total number
of topplings after the addition of a slope unit.
The avalanche size probability density function, $P(s;L)$,
in a system of size $L$ follows simple scaling above a lower cutoff $s_{\ell}$,
\begin{equation}
P(s;L) = a s^{-\tau} \GC\left(\frac{s}{b L^D}\right) \text{ for } s>s_{\ell},
\elabel{simple_scaling}
\end{equation}
where $D$ is the gap exponent (avalanche dimension) and $\tau$ the
avalanche size exponent.  The two constants $a$ and $b$ are metric
factors. The scaling function $\GC(x)$ falls off sufficiently fast such
that all moments of the avalanche size probability density function exist in a finite system.
For $n > \tau-1$, the leading order of the $n$th moment is \cite{DeMenechStellaTebaldi:1998}
\begin{equation} \elabel{moment_scaling}
\ave{s^n}(L) = \int_0^\infty\!\! s^n P(s;L)ds = a (b L^D)^{n+1-\tau} \gc_n + \cdots,
\end{equation}
where $\gc_n$ depends only on the scaling function $\GC$
\cite{JensenPruessner:2003}. The sub-leading terms
represent, for example, corrections to scaling \cite{Wegner:1972} and
the presence of a lower cutoff.

The numerical estimates of the exponents for the one-dimensional SOC Oslo
model are very well established with $D=2.25(2)$ and $\tau=1.555(2)$
\cite{ChristensenETAL:1996,PaczuskiBoettcher:1996,JensenPruessner:2003,Christensen:2004},
independent of the exact boundary condition at $i = L$
\footnote{The simulations in Ref. \cite{JensenPruessner:2003} refer to
a boundary condition where site $i=L$ relaxes according to $z_L\to z_L-2$ and $z_{L-1}\to z_{L-1}+1$.}.
In the following, we address the key questions: Is it possible to recover these
exponents, characterizing the avalanche behavior and therefore the SOC
aspect of the model, in its AS counterpart?  Is the critical slope density
identical in both models?

\prepsection{The AS Oslo model} The recipe mentioned in the introduction can be
``inverted'' to transform the SOC Oslo model into an AS version by imposing
periodic boundary conditions such that sites $i=1$
and $i=L$ are nearest neighbors, that is, for all sites toppling
(including $i=1$ and $i=L$), one slope unit is moved to each of the two
nearest neighbors.  The resulting model is translationally invariant, so
that the original interpretation of the slopes as height differences
between columns of rice breaks down \cite{Bak87,ChristensenETAL:1996}.
The slope density
increases in steps of $1/L$ when the system is driven.
If the external drive triggers
activity, the avalanche will propagate until the systems falls into an
absorbing state again.
Contrary to the SOC Oslo model, there is no dissipation
mechanism coupled to the activity
to decrease the slope density in the AS Oslo model.

Every finite system will eventually fall into an absorbing state, provided
that $\zeta \le 2$.  However, numerically it becomes clear that for
system sizes $L \gg 1$ there exists a critical slope density,
$\zeta_c < 2$, above which the absorbing states become practically
inaccessible. Defining the activity as the density of sites where
$z_i>z_i^c$, the activity picks up sharply at the critical slope density,
$\zeta_c$. 
However, one fundamental problem with a definition of an
instantaneous activity is the Abelian nature of the Oslo model \cite{Dhar:2003},
meaning that the order in which sites are relaxed is irrelevant. Without an 
{\it a priori} order of relaxation, there can be no {\it a priori}
microscopic timescale, and the
temporal behavior of the instantaneous activity depends on the
choice of microscopic timescale.
It will be argued below that $\zeta_c$
and all relevant observables can be be obtained while avoiding these
ambiguities.

\prepsection{Simulation procedure} Starting from an empty configuration,
$\zeta = 0$, slope units are added at site $i=1$ although, of course,
all sites are equivalent because of translational invariance.  Just as
in the SOC Oslo model, slope units are added only when activity has
ceased. Thus, one obtains avalanche sizes for slope densities increasing in
steps of $1/L$.  At small slope densities, only small avalanches occur and until
avalanches wrap around the system, the system size cannot have any
effect on the dynamics.  In this regime, we find that the behavior of
the model essentially resembles that of the one dimensional BTW model \cite{Bak87}
where the avalanche size scales like the square of the number of
slope units added, $(\zeta L)^2$.

At sufficiently large slope densities, the avalanche sizes start to
deviate from the above behavior. According to the original arguments
\cite{Vespignani97}, one expects that at the AS critical point,
$\zeta_c$, the avalanche size probability density function follows simple scaling,
\Eref{simple_scaling}.
Let $\ave{s^n}(\zeta; L)$ denote the $n$th
moment of the avalanche size probability density function in the AS Oslo model of size $L$ at
slope density $\zeta$. Ignoring corrections to
scaling and the existence of a lower cutoff in \Eref{moment_scaling},
one expects that at $\zeta_c$ there exist exponents
\begin{equation}
\gamma_n = D (n+1-\tau)
\elabel{scaling_relation}
\end{equation}
such that
\begin{equation} \elabel{crossing}
\frac{\ave{s^n}(\zeta_c; L)}{L^{\gamma_n}}= a b^{n+1-\tau} \gc_n \quad \text{for $L \gg 1$}.
\end{equation}
Note that the right hand side is independent of $L$ but depends on $n$, the order of the moment.
However, away from $\zeta_c$ the ratio on the left hand
side depends on $L$. 
Therefore, for all $n$ and three distinct system sizes $L_1,L_2,L_3$, there exist a unique
slope density, $\zeta_c$, and exponent, $\gamma_n$, such that
\begin{equation} \elabel{crossing_pair}
\frac{\ave{s^n}(\zeta_c; L_1)}{L_1^{\gamma_n}}=
\frac{\ave{s^n}(\zeta_c; L_2)}{L_2^{\gamma_n}} = \frac{\ave{s^n}(\zeta_c; L_3)}{L_3^{\gamma_n}}.
\end{equation}
Graphically, $\zeta_c$ and $\gamma_n$ are determined by plotting, for
each system size, the rescaled $n$th moment, $\ave{s^n}(\zeta;
L)/L^{\gamma_n}$, versus the slope density $\zeta$ and adjusting the exponent
$\gamma_n$ until the graphs intersect at a single point at $\zeta_c$,
see \Fref{Fig1}.  The slope of the graphs of the resulting exponents,
$\gamma_n$, versus the order of the moment, $n$, determines the gap
exponent, $D$, while the intersection with the $n$-axis gives $\tau-1$,
see \Eref{scaling_relation}.

Since the moments are measured for slope densities increasing in steps
of $1/L$, the location of the crossings requires interpolation between
the data points.  This introduces arbitrariness, which can, however, be
reduced strongly by choosing system sizes which are commensurable with
an estimated critical slope density such that $\zeta_c L \in
\mathbb{N}$. In a preliminary simulation with $L = 1024, 2048, 4096,
8192$ we estimated $\zeta_c = 1.73265$ which suggests $L = 1268, 2536,
5072, 10144$ as suitable system sizes. The numerical estimates presented
below refer to an exponential interpolation scheme but they are very
similar to those obtained from linear interpolation.  All systems were
initialized with $\zeta=0$ and gradually filled to slightly above
$\zeta_c$, where the average avalanche size becomes extremely large. In
each simulation, a configuration slightly below $\zeta_c$ was recorded
and the model driven from this configuration $100$ times to improve
statistics. For every system size, we average over approximately $2000$
\emph{independent} realizations starting from $\zeta = 0$. Thus there is
a total of approximately $200,000$ \emph{correlated} realizations in the
neighborhood of $\zeta_c$. The error bars reported below are based only
on the $2000$ \emph{independent} realizations, thereby grossly
overestimating the errors.

\begin{figure}[h]
\begin{pspicture}(0,0)(5,6)
\rput(2.75,3){\scalebox{0.35}{\includegraphics{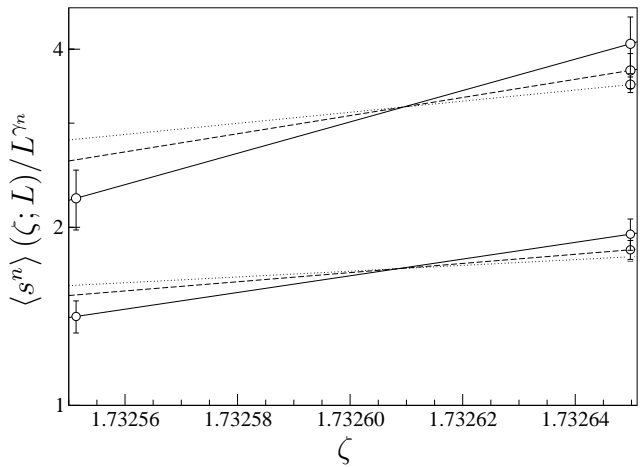}}}
\fontsize{12pt}{12pt}
\rput(2.75,-0.1){$\zeta$}
\rput{90}(-1.5,3.1){$\ave{s^n}(\zeta;L)/L^{{\gamma}_n}$}
\end{pspicture}
\caption{The logarithm of the rescaled moments $\ave{s^n}(\zeta;L)/L^{\gamma_n}$ versus
the slope density $\zeta$ for $n = 1,2$ (open circles). The exponential interpolations between the data points for
increasing system sizes are marked with lines of increasing dash length.
For the triplet $L = 2536, 5072, 10144$, the lower set of graphs for $n=1$ intersect in a single
point at $\zeta_c = 1.732608$ with $\gamma_1 = 2.064$ and the upper set of graphs for $n=2$
intersect at $\zeta_c = 1.732609$ with $\gamma_2 = 4.342$.
} \label{F:Fig1}
\end{figure}

\prepsection{Results: critical slope density} The numerical estimate for the
critical slope density, $\zeta_c$, is determined as the slope density at which the
graphs for different system sizes cross for the rescaled $n$th moment,
see Fig.~\ref{F:Fig1}.

In principle, each triplet of system sizes
produces an estimate of $\zeta_c$ for each moment.
For example, from the crossings of the triplet $L=2536,5072,10144$ we
find $\zeta_c=1.73261(5)$ for $n=1$ and $\zeta_c=1.73261(5)$ for $n=2$,
see Fig.~\ref{F:Fig1}.
For the triplet $L=1268,2536,5072$ the resulting estimates
are $\zeta_c=1.73262(8)$ for $n=1$ and
$\zeta_c=1.73259(7)$ for $n=2$.
Ignoring errors,
all triplets and moments yield
\begin{equation}
1.73257 \leq \zeta_c \leq 1.73262 \ .
\end{equation}
There is no established systematic way to determine the value of
$\zeta_c$ in the limit $L \to \infty$ but from the small change in the
estimate as the system sizes are increased, it seems reasonable to
estimate the critical slope density by averaging over all moments
$n = 1,2,3,4$ and all triplets yielding $\zeta_c=1.73260(2)$.

\begin{figure}[h]
\begin{pspicture}(0,0)(5,6)
\rput(2.75,3){\scalebox{0.35}{\includegraphics*{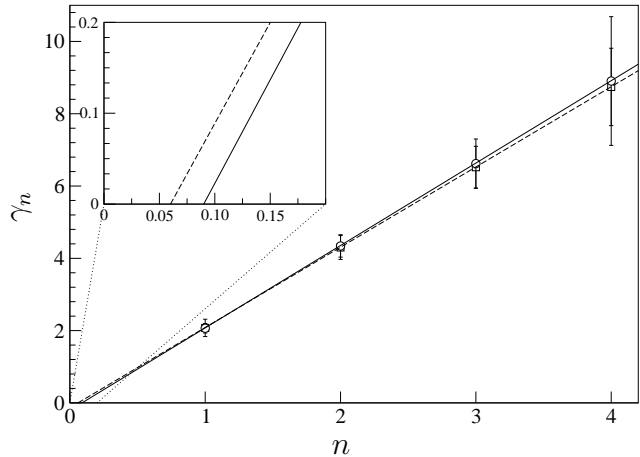}}}
\fontsize{12pt}{12pt}
\rput(2.75,-0.1){$n$}
\rput{90}(-1.5,3.1){$\gamma_n$}
\end{pspicture}
\caption{The estimated exponents $\gamma_n$ versus the order, $n$, of
the moment for the triplets $L=1268,2536,5072$ (open squares) and
$L=2536,5072,10144$ (open circles).  Assuming simple finite-size
scaling it follows from \Eref{scaling_relation} that the slope of the
graphs is the gap exponent, $D$, while the intersection with the
$n$-axis is $\tau-1$, see inset (dotted lines indicate magnified
region). Linear regression yields $D = 2.22(23), \tau= 1.07(18)$ for the
first triplet (dashed line) and $D=2.28(23), \tau =1.09(13)$ for the
second triplet (solid line).} \label{F:Fig2}
\end{figure}

\prepsection{Results: critical exponents}
Since crossings rely on an appropriate choice of $\gamma_n$, estimates
of the critical slope density $\zeta_c$ and $\gamma_n$ go hand in hand,
see Fig.~\ref{F:Fig1}.  Plotting $\gamma_n$ versus the order of the
moment, $n$, produces according to \Eref{scaling_relation} an estimate
for $D$ and $\tau$. This procedure can be
performed for every triplet of system sizes. For $L=1268,2536,5072$ we
find $D = 2.22(23)$ and $\tau=1.07(18)$ while for $L=2536,5072,10144$ the
resulting exponents are $D = 2.28(23)$ and $\tau=1.09(13)$, see Fig.~\ref{F:Fig2}.

The errors of
these exponents were estimated by standard error propagation.
Just as for the estimate of $\zeta_c$, the procedure only provides
exponents for a particular set of system sizes. However, even the
preliminary simulations  with $L=1024, 2048, 4096, 8192$ are fully compatible with
the exponents $D$ and $\tau$ reported above. Based on all four possible
triplets of system sizes, it therefore seems justified to estimate 
$D=2.25(8)$ and $\tau=1.08(5)$, in the limit $L \to \infty$.

\prepsection{Discussion}
Table \ref{Table} summarizes the numerical estimates for the
critical slope densities and exponents for the AS and boundary driven 
SOC Oslo model.
\begin{table}[h]
\begin{tabular} {|l|c|c|c|c|}
\hline
                      & \multicolumn{3}{c}{\phantom{22}Quantity} &                                               \\ \hline
Oslo model            &   $\zeta_c$              &  $D$         &  $\tau$               & \phantom{1}$\gamma_1$\phantom{1}  \\ \hline
AS                    &   $1.73260(2)$           &  $2.25(8)$   &  $1.08(5)\phantom{2}$ & $2.07(6)$            \\ \hline
SOC, boundary driven  &   $1.7326(3)\phantom{0}$ &  $2.25(2)$   &  $1.555(2)$           &  $1$        \\ \hline
SOC, bulk driven      &   $1.734(2)\phantom{66} $            &  $2.25(3)$   &  $1.10(3)\phantom{2}$ &  $2$         \\ \hline
\end{tabular}
\caption{The critical slope density $\zeta_c$, the gap exponent $D$, the avalanche size exponent $\tau$, and
the exponent $\gamma_1$ in the AS Oslo model (all errors based on
 averaging over the four triplets), SOC Oslo model driven at the boundary
\cite{ChristensenETAL:1996,PaczuskiBoettcher:1996,JensenPruessner:2003,Christensen:2004}
or in the bulk \cite{Sorenssen99}.
For SOC Oslo models, a simple conservation argument in the stationary regime
determines $\gamma_1$.
}\label{Table}
\end{table}

We are not aware of a systematic study of the \emph{average slope
density} $\ave{\zeta}(L)$ in the SOC Oslo model. However, results from
simulations of systems sizes $L=1024, 2048, 4096, 8192$ are consistent
with $\zeta_c - \ave{\zeta}(L)\propto L^{-x}$ with $x\approx 0.7$ and
$\zeta_c=1.7326(3)$, where the error bar is based on visual
inspection. Similarly for the bulk driven model (see below) we find
$\zeta_c=1.734(2)$. 

It is surprising how well the \emph{non-universal} slope density of the
SOC Oslo model is reproduced by the AS model.
This, however, is exactly what is predicted by
the simple mechanism put forward by Vespignani and Zapperi \cite{Vespignani97}.

The avalanche size exponent $\tau =1.08(5)$ in the AS Oslo model is
inconsistent with the value $\tau = 1.555(2)$ in the SOC Oslo
model. However, the gap exponent $D =2.25(8)$ in the AS Oslo model is
consistent with the value $D = 2.25(2)$ reported in the literature for
the SOC Oslo model
\cite{ChristensenETAL:1996,PaczuskiBoettcher:1996,JensenPruessner:2003,Christensen:2004}.
While the former result questions the one-to-one correspondence between
the SOC and AS Oslo model, one might understand this
discrepancy as follows: The exponent actually characterizing the Oslo
model is the gap exponent $D$. The gap exponent is deeply rooted in
the model and is also present in the corresponding field theory
\cite{PaczuskiBoettcher:1996,Pruessner:2003}. In contrast, in the SOC
Oslo model, the avalanche size exponent $\tau$ is determined by the gap
exponent $D$ via the scaling relation \Eref{scaling_relation}, since $\gamma_1$ can be derived
by simple conservation arguments in the stationary regime; driving the SOC Oslo model at any
fixed, \emph{absolute} position, such as $i=1$ in the boundary driven
SOC Oslo model described above, leads to $\gamma_1 = 1$, implying
$D(2-\tau)=1$. Driving the model randomly in the bulk with slope units
or at any \emph{relative} position, such as $i=L/2$, leads to $\gamma_1
= 2$, implying $D(2-\tau)=2$. The latter scaling law produces
$\tau=1.111(8)$ from $D=2.25(2)$, well compatible with the result found
in the AS Oslo model.  In fact, the exponents $D=2.25(3)$ and
$\tau=1.10(3)$ have been reported in the literature for a variant of the
bulk-driven Oslo model \cite{Sorenssen99}, but interpreted as evidence
against the bulk-driven SOC Oslo model being in the same universality
class as the boundary driven SOC Oslo model.

Contrary to the SOC Oslo models,
there exists no conservation argument to
determine the value of $\gamma_1$ in the AS Oslo model.
However, because of the periodic boundaries, the AS Oslo model seems to
be related to the bulk-driven SOC Oslo model rather than the
original boundary-driven SOC Oslo model. 
Therefore, the discrepancy in the
avalanche size exponent, $\tau$, in the AS Oslo model with respect to
the SOC Oslo model driven at any fixed position might very well be caused by the
difference in the external drive and the resulting scaling relations. In
fact, the AS Oslo model analyzed above seems to be a perfect AS version
of an SOC Oslo model driven in the bulk at, say, $i=L/2$.

Since the SOC Oslo model is equivalent to an interface depinning model
\cite{PaczuskiBoettcher:1996,Pruessner:2003},
the latter should also have an equivalent AS version, see discussion in
Ref. \cite{PruessnerPHDThesis:2004}.
It is also
conjectured that the train model is in the same universality class as
the Oslo model \cite{PaczuskiBoettcher:1996}.
Whether imposing periodic boundary conditions
on the train model will transform it into its AS analogue is an
intriguing question which is beyond the scope of this brief report.

In conclusion, we have transformed the SOC Oslo model into an AS Oslo model.
We have numerically determined the critical slope densities and obtained numerical
estimates for the gap exponent, $D$, and avalanche size exponent, $\tau$,
characterizing the avalanche behavior assuming simple finite-size scaling.
The critical slope densities and the gap exponents are identical in the two
models. However, the avalanche size exponents are different.
The question remains whether one can construct a
``proper'' AS Oslo model, corresponding to the boundary driven SOC Oslo
model. 

\prepsection{acknowledgments}
The authors wish to thank 
A. Thomas as well as D. Moore, B. Maguire and P. Mayers 
for their
technical support of the SCAN cluster at the Department of Mathematics at
Imperial College London. 
The authors are indebted to PGP, University of Oslo, Norway for
 support and hospitality during their visit. 
NRM is very grateful to the Beit Fellowship.
OP and GP gratefully acknowledge the support of EPSRC, GP also acknowledges
the support of the AvH Foundation and the NSF.

\bibliography{periodic_oslo}
\end{document}